# MoS$_2$ P-type Transistors and Diodes Enabled by High Workfunction MoO$_x$ Contacts

Steven Chuang[1,2,3], Corsin Battaglia[1,2,3], Angelica Azcatl[4], Stephen McDonnell[4], Jeong Seuk Kang[1,2,3], Xingtian Yin[1,2,3], Mahmut Tosun[1,2,3], Rehan Kapadia[1,2,3], Hui Fang[1,2,3], Robert M. Wallace[4], Ali Javey[1,2,3,*]

[1]Electrical Engineering and Computer Sciences, University of California, Berkeley, CA, 94720

[2]Materials Sciences Division, Lawrence Berkeley National Laboratory, Berkeley, CA 94720

[3]Berkeley Sensor and Actuator Center, University of California, Berkeley, CA, 94720

[4]Department of Materials Science and Engineering, The University of Texas at Dallas, Richardson, TX, 75080

* Electronic mail: ajavey@berkeley.edu

*Abstract-* The development of low-resistance source/drain contacts to transition metal dichalcogenides (TMDCs) is crucial for the realization of high-performance logic components. In particular, efficient hole contacts are required for the fabrication of *p*-type transistors with MoS$_2$, a model TMDC. Previous studies have shown that the Fermi level of elemental metals is pinned close to the conduction band of MoS$_2$, thus resulting in large Schottky barrier heights for holes with limited hole injection from the contacts. Here, we show that substoichiometric molybdenum trioxide (MoO$_x$, x<3), a high workfunction material, acts as an efficient hole injection layer to MoS$_2$ and WSe$_2$. In particular, we demonstrate MoS$_2$ *p*-type field-effect transistors and diodes by using MoO$_x$ contacts. We also show drastic on-current improvement for *p*-type WSe$_2$ FETs with MoO$_x$ contacts over devices made with Pd contacts, which is the prototypical metal used for hole injection. The work presents an important advance in contact engineering of TMDCs and will enable future exploration of their performance limits and intrinsic transport properties.

Transition metal dichalcogenides (TMDCs) offer ultra-thin, uniform channel thicknesses for unparalleled gate control, and are a strong candidate for future electronics [1] [2] [3] [4] [5]. In order to apply TMDCs to low-power, high-performance complementary logic applications, both *n*- and *p*-type field effect transistors (NFETs and PFETs) must be developed. The polarity of a FET is determined by the type of charge carriers that can be injected from the source contact into the semiconductor channel. In a conventional metal-oxide-semiconductor FET (MOSFET), this is achieved by heavily doping the source/drain contacts to either *p+* or *n+* for *p* and *n*-type transistors respectively. Similarly, in a Schottky MOSFET, where metal contacts are directly fabricated on the semiconductor, the device polarity is determined by the Schottky barrier (SB) heights for electrons and holes at the source contact. A small SB height to the conduction or valence band leads to *n* or *p*-type FETs, respectively. SB heights, in principle, can be controlled by the work function potential of metal contacts. To date, most reported TMDC FETs have been based on the Schottky device architecture given its ease of fabrication. While TMDC NFETs have been relatively well studied [3] [4] [5] [6], there has been difficulty fabricating high-performance TMDC PFETs, largely limited by hole injection at the source/drain (S/D) contacts due to large SB heights to the valence band. Traditionally, the high work function metal palladium (Pd) has been used as the most popular contact material to the valence band of various nanostructures, including nanotubes, graphene, and organics [7] [8] [9] [10]. However Pd alone has proven insufficient as a hole contact for TMDC devices. With a workfunction of 5.1 eV [11], the Fermi level of ultra-clean Pd lies slightly above the valence band maximum of $MoS_2$ (Fig. 1a) [12] [13]. However, most previously reported Pd-contacted $MoS_2$ devices exhibit *n*-type behavior with high contact resistance instead of *p*-type behavior, which is commonly ascribed to Fermi-level pinning at the $MoS_2$ contact interface [6] [14]. A recent study has shown that limited hole injection can be observed in Pd-contacted $MoS_2$ devices, but only in the limit of large gate fields when the SBs are sufficient thinned by the electrostatic fields [15]. On the other hand Pd-contacted $WSe_2$ PFETs show high contact resistances and require surface charge transfer doping to thin the SBs and allow tunneling of holes at the contacts [2].

Here we explore substoichiometric molybdenum trioxide ($MoO_x$, x<3) as a promising material for hole injection into TMDCs without doping the semiconductor body. $MoO_x$ exhibits a high work function potential of up to ~6.6 eV (see Fig. 1a) [16] exceeding those of elemental metals [11]. While $MoO_x$ has been previously used as hole contacts in organic electronics [17] [18], its application to inorganic semiconductors was extended only recently [16] [19]. Here we demonstrate a series of TMDC devices with $MoO_x$ contacts that highlight unambiguously the advantages of $MoO_x$ hole contacts over conventionally explored elemental metal contacts. $MoS_2$ FETs with $MoO_x$ contacts present *p*-type behavior despite the notorious Fermi level pinning to the conduction band previously observed [6]. $MoS_2$ Schottky diodes with asymmetric $MoO_x$ and Ni contacts clearly display rectifying behavior. Finally, $WSe_2$ PFETs with $MoO_x$ contacts show an order of magnitude increase in on-current when compared to Pd-contacted $WSe_2$ PFETs.

Fig. 1b compares monochromatic x-ray photoelectron spectra (XPS) of the valence band region of $MoO_x$ and Pd films [20]. While Pd shows a strong photoelectron signal below the Fermi energy ($E_F$) with a clear metallic Fermi-Dirac step centered at $E_F$, the valence band of thermally evaporated $MoO_x$ possesses a weak characteristic defect band in the band gap derived from oxygen vacancies, whose tail reaches all the way up to $E_F$. Consequently $MoO_x$ can be classified as a semiconducting oxide with a metallic defect band. Its workfunction can exceed 6.6 eV, but is known to strongly depend on carbon contamination [16]. For practical applications, $MoO_x$ can thus be considered to act as a high workfunction metal with a low density of states at the Fermi level. Consequently most metals should form ohmic contacts with $MoO_x$.

In order to confirm ohmic contact between $MoO_x$ and Pd, the current-voltage characteristics across Pd/$MoO_x$/Pd stacks were measured. Stacks of 20nm Pd/ $MoO_x$/ 40nm Pd were fabricated by photolithography, evaporation and lift-off. The $MoO_x$ thickness was varied from 100 nm to 400 nm. Fig. 1c shows the clearly linear current-voltage characteristics of the stacks which confirm ohmic behavior between $MoO_x$ and Pd. Fig. 1d shows the total resistance of these devices as a function of $MoO_x$ thickness. The resistance of a single contact is extracted from half the y-intercept of the linear fit as ~200 $\mu\Omega\bullet cm^2$. The resistivity extracted from the slope of the linear fit of the plot is ~200 $\Omega\bullet cm$. Although this resistivity is high, keeping the $MoO_x$ layer thin enough (i.e., sub-50 nm) guarantees efficient carrier transport.

We now turn to the fabrication of $MoS_2$ PFETs with $MoO_x$ contacts. $MoS_2$ flakes were first exfoliated mechanically onto a 260 nm $SiO_2$/Si substrate. A 1 hour acetone bath was used to clean any organic residues from the chip after exfoliation. Symmetrical 30 nm $MoO_x$/ 30 nm Pd contacts were defined on the $MoS_2$ flakes via photolithography, evaporation and lift-off. The channel length between the contacts is ~7 μm. In order to minimize workfunction lowering due to carbon contamination of $MoO_x$, several precautions were taken. Thermal evaporation of $MoO_x$ was carried out after ~12 hours of pumping at a base pressure of ~$8\times10^{-7}$ Torr at a rate of 0.5 Å/s. $MoO_3$ powder (99.9995% purity, Alfa Aesar) was used as the $MoO_x$ evaporation source throughout this study. Electron-beam evaporation of Pd was performed right after $MoO_x$ deposition without breaking vacuum.

A schematic and optical microscope image of a representative $MoS_2$ PFET with $MoO_x$ contacts are shown in figure 2a. Corresponding $I_{ds}$ vs $V_{gs}$ characteristics are shown in figure 2b. All TMDC devices in this study were measured in vacuum in order to isolate effects from exposure to ambient, such as the adsorption of oxygen and water [4]. The thickness of the $MoS_2$ flake was measured as 40 nm with atomic force microscopy (AFM). Clear *p*-type characteristics with $I_{on}/I_{off}$~$10^4$ are obtained, indicating hole contact to the valence band. In contrast, control devices fabricated with Pd contacts (without $MoO_x$) exhibit clear *n*-type characteristics (see Supplementary Information, Fig. S1a), consistent with literature on mechanically exfoliated $MoS_2$ [6]. 2D simulations coupling drift-diffusion and Poisson relations were performed with TCAD Sentaurus to extract the SB heights from the experimental $I_{ds}$ vs $V_{gs}$ results. An in-plane effective mass of 0.45 $m_0$ for electrons and 0.43 $m_0$ for holes were assumed [21]. An electron mobility of 200 $cm^2$/V·s and hole mobility of 86 $cm^2$/V·s were assumed [22] [5]. The subthreshold slope (SS) of 410 mV/dec was fit with a uniform density of interface traps $D_{it}$ of $6\times10^{11}$ $cm^{-2}eV^{-1}$ across the $MoS_2$ bandgap at the $MoS_2/SiO_2$ interface. This value of SS is reasonable given we have multi-layer flakes on a thick (260 nm) backgate oxide. Threshold voltage shifts were applied to match each simulated curve with its respective experimental data. From the qualitative band diagram in figure 2d it is evident that with the non-negligible barrier to the valence band, we expect tunneling and thermionic emission to dominate the on-current characteristics. Thus a nonlocal tunneling model based on the Wentzel-Kramers-Brillouin (WKB) formalism was implemented at the contacts. An out-of-plane effective mass of 1.0 $m_0$ was used as the hole tunneling mass [21]. A hole SB height of 0.31 eV was used to fit the on-current to the experimental results. Such a low barrier height is surprising given that elemental metals have been shown to be Fermi level pinned ~1.1-1.2eV from the valence band of $MoS_2$ [14]. A good fit is obtained to the $I_{ds}$ vs $V_{gs}$ curve in the subthreshold, linear and saturation regimes of the device as shown in figure 2b. $I_{ds}$ vs $V_{ds}$ characteristics are shown in figure 2c. Clear linear and saturation regimes are exhibited, indicating standard MOSFET device operation. The saturation current of typical long channel FETs is proportional to $(V_g-V_t)^2$, however in our device we observe that saturation current is proportional to $V_g-V_t$ instead. This observation suggests that the device has non-negligible series resistance, most likely from the SB barrier height at the contacts [23].

Next we investigate the origin of the hole injection improvement in $MoO_x$ contacts as compared to elemental metals. An interface Fermi-level pinning parameter $S = \frac{\partial \Phi_p}{\partial \psi_c} = -\frac{\partial \Phi_n}{\partial \psi_c} = -0.1$ ($\Phi_{n/p}$ =electron/hole SB height, $\psi_c$ = contact workfunction) was previously extracted for elemental metal contacts [14]. Using this pinning parameter and the highest work function $\psi_c = $ 6.6 eV we observed for $MoO_x$, we expect a lower bound SB height of $\Phi_p$~ 1 eV for $MoO_x/MoS_2$ contacts. This value is significantly larger than our experimental observations and suggests a lower degree of Fermi-level pinning at $MoO_x/MoS_2$ contacts as compared to elemental contacts. This may be expected given the difference in the nature of the interface chemical bonding and the density of states at the Fermi level for $MoO_x$. Specifically, due to the low density of states at the Fermi level (see again Fig. 1b) and the localized nature of the defect states in $MoO_x$, its tendency to form metal induced gap states is possibly less pronounced than that of elemental metals such as Pd [24] [25]. Alternatively interface states could originate from native defects of the $MoS_2$ surface or from surface damage caused by metal evaporation. If so, MoO$_x$ possibly passivates and reduces the number of such states. Further experimental and theoretical investigations are necessary to understand the contact/TMDC interface for both MoO$_x$ and elemental metals. Nevertheless, the work here clearly suggests that the advantage of MoO$_x$ contacts for hole injection is not only due to its high work function, but also due to its better interface properties (i.e., lower degree of interface Fermi-level pinning) with TMDCs.

MoS$_2$ Schottky diodes were studied in order to further demonstrate the utility of MoO$_x$ as an effective hole contact to MoS$_2$. The process flow was identical to that for the MoS$_2$ PFETs, other than the fact that two photolithography steps were used to pattern Ni and MoO$_x$/Pd asymmetric contacts to the same MoS$_2$ flake. A device schematic and qualitative band diagram are shown in figures 3a and b. Specifically, Ni is used as an electron contact with a small SB height ($\Phi_{n,Ni}$) to the conduction band of MoS$_2$ according to previous reports [3] [6] and our control experiments reported in Fig. S1b. On the other hand, MoO$_x$ is used as the hole contact with a small SB height ($\Phi_{p,MoO_x}$) to the valence band of MoS$_2$ as previously discussed. The resulting electrical measurements with a grounded Si substrate for a 24 nm thick MoS$_2$ flake are shown in figure 3c. Clear rectification is shown with a forward/reverse bias current ratio of up to ~$10^5$. The direction of the rectification is consistent with that originating from the two asymmetric contacts discussed above (see band diagram in Fig. 3b). An ideality factor n of 1.4 at room temperature is extracted from the ideal diode region. The ideality factor of a diode typically varies between 1 and 2 depending on the relative contribution of current from diffusion and recombination, respectively, assuming mid-gap trap states. The low value extracted for the ideality factor indicates a low contribution of recombination current and a low density of trap states at the MoO$_x$/MoS$_2$ junction and in the unintentionally doped MoS$_2$ [26]. By plotting the natural log of the reverse bias current $I_{rev}$ at $V_{sd}$ = -2 V as a function of 1/kT, the activation energy $E_A$ of the reverse bias was extracted as 0.34±0.02 eV (Fig. 3d), which most likely corresponds to phonon assisted tunneling mechanisms commonly observed in reverse biased Schottky diodes [27]. A small temperature dependence is observed in the $V_{sd}$ >1 V forward biased region of the diode. Assuming that this region is dominated by series resistance from the $MoS_2$ flake, this observation can be attributed to the small temperature dependence of the shallow dopants (i.e., un-intentional impurities) in $MoS_2$ in this temperature range.

Given the success in contacting the valence band of $MoS_2$, we next fabricated PFETs using $WSe_2$, a promising *p*-type TMDC [2]. Unlike $MoS_2$, $WSe_2$ has been shown to exhibit a surface Fermi level pinning closer to the valence band edge, thereby, making it easier to obtain PFETs by using various metals. The fabrication process was analogous to the $MoS_2$ PFET other than the different flake exfoliated. A schematic of the device is shown in figure 4a. $WSe_2$ devices with 30 nm $MoO_x$/ 30 nm Pd contacts as well as a reference device with 30 nm Pd contacts were fabricated and compared to each other. The thicknesses of the $WSe_2$ flakes were 32 nm and 29 nm for the $MoO_x$ and Pd contacted flakes, respectively. The resulting $I_d$ vs $V_{gs}$ characteristics measured in vacuum are shown in figure 4b. More than one order of magnitude improvement in ON current is observed in the $MoO_x$ contacted $WSe_2$ device compared to the Pd contacted device. Sentaurus simulations were performed to investigate the origin of this improvement. A uniform $D_{it}$ of $1.2 \times 10^{12}$ $cm^{-2}eV^{-1}$ across the $WSe_2$ bandgap was used to fit the SS of 970 mV/dec. We assumed 0.3 $m_0$ for the in-plane effective mass of holes $m_h$ [28]. Given the $WSe_2$ reduced electron-hole mass $m_r$ = 0.24 $m_0$, the effective mass of electrons was assumed to be $m_e = (\frac{1}{m_r} - \frac{1}{m_h})^{-1} = 1.2$ $m_0$ [29]. An electron mobility of 200 $cm^2/V \cdot s$ and hole mobility of 329 $cm^2/V \cdot s$ were used [30] [31]. Again a non-local WKB model was used to simulate the contacts, with a 0.9 $m_0$ out-of-plane effective mass used as the hole tunneling mass parameter [32]. Hole SB heights of 0.29 eV and 0.37 eV were used to fit the on current of the $MoO_x$ and Pd contacted devices, respectively. From the qualitative band diagrams in figure 4c, we see the lower hole barrier height in the $MoO_x$ contacted devices facilitates improved hole injection, resulting in higher on currents. The simulated curves match the experimental results well. The overestimation of the simulation current for the Pd contacted device in the subthreshold region could be ascribed to the oversimplification of the $WSe_2$/contact and dielectric interfaces. Atomic simulations are needed in the future to better account for the $WSe_2$/contact interfaces.

We also characterized the stability of devices in air. The PFETs measured are highly stable over time, showing minimal change in IV characteristics over the course of >2 weeks exposure to air (Supporting Information, Fig. S1a). However when measured in air instead of vacuum, all $MoO_x$ devices show a reversible lowering of on-current (Supporting Information, Fig. S1b). Original device characteristics are restored upon placement in vacuum. This observation can be attributed to the sensitivity of the $MoO_x$ work function to ambient gas exposure [16]. This behavior is similar to elemental metal contacts to devices, which also show barrier height modulation due to gas exposure, and can be remedied by encapsulating the device [7] [9].

In conclusion, this study explores high work function $MoO_x$ contacts to the valence band of TMDCs for efficient hole injection, addressing a key challenge for obtaining high performance *p*-type and complementary logic components. $MoO_x$ contacts to $MoS_2$ enables fabrication of PFETs with $I_{on}/I_{off}\sim10^4$ despite previous studies showing metals being Fermi level pinned near the conduction band edge of $MoS_2$ [14]. $MoS_2$ Schottky diodes with asymmetric $MoO_x$ and Ni contacts exhibit rectifying behavior. Finally, $WSe_2$ PFETs with $MoO_x$ contacts exhibit an order of magnitude improvement in $I_{on}$ over Pd contacted $WSe_2$ PFETs. The observed FET behavior could be captured well by 2D simulations. Overall this study is an invitation to explore transition metal oxides with extreme work functions as selective carrier contacts to TMDCs for realization of high performance devices.

**Acknowledgements:** The device fabrication and characterization portion of this work was funded by the Director, Office of Science, Office of Basic Energy Sciences, Material Sciences and Engineering Division of the U.S. Department of Energy under Contract No. DE-AC02-05CH11231. XPS analysis and materials characterization portion of this work was funded by the Center for Low Energy Systems Technology (LEAST), one of six centers supported by the STARnet phase of the Focus Center Research Program (FCRP), a Semiconductor Research Corporation program sponsored by MARCO and DARPA.

**Supporting Information Available:** Reference $MoS_2$ FETs with Pd and Ni contacts exhibiting *n*-type behavior. Air stability of $MoS_2$ PFETs with $MoO_x$ contacts. This material is available free of charge *via* the internet at http://pubs.acs.org.

# Figure Captions

**Figure 1**. (a) Valence and conduction band positions with respect to vacuum level for $MoS_2$, $WSe_2$, Pd and $MoO_x$. (b) Valence band photoelectron spectra for $MoO_x$ and Pd films evaporated in ultra-high vacuum conditions using monochromatized Al $K_\alpha$ radiation. (c) Current-voltage characteristics across Pd/$MoO_x$/Pd stacks indicating good Ohmic contact with a schematic of the test structure in the inset. (d) Resistance of Pd/$MoO_x$/Pd stacks as a function of $MoO_x$ thickness.

**Figure 2**. (a) Schematic and optical microscope image, (b) $I_{ds}$ vs $V_{gs}$ and (c) $I_{ds}$ vs $V_{ds}$ characteristics for a representative $MoS_2$ PFET with $MoO_x$ contacts. (d) Qualitative band diagrams for the ON (top panel) and OFF (bottom panel) states of the $MoS_2$ PFET.

**Figure 3**. (a) Schematic for a representative $MoS_2$ Schottky diode made with asymmetric metal contacts. (b) Qualitative band structure of the device with asymmetric Ni and $MoO_x$ electrodes used as electron and hole contacts, respectively. (c) Temperature dependent $I_d$ vs $V_{sd}$ electrical characteristics of the diode. (d) Barrier height extraction of the reverse bias current ($I_{rev}$) at $V_{sd} = -2V$ from temperature dependent measurements.

**Figure 4**. (a) Schematic, (b) $I_{ds}$ vs $V_{gs}$ characteristics and (c) qualitative band diagrams for $WSe_2$ devices contacted with $MoO_x$ (left panel) and Pd alone (right panel). The hole barrier heights are indicated as $\Phi_{p,MoO_x}$ and $\Phi_{p,Pd}$ for the $MoO_x$ and Pd contacted devices, respectively.

**Figure 1**

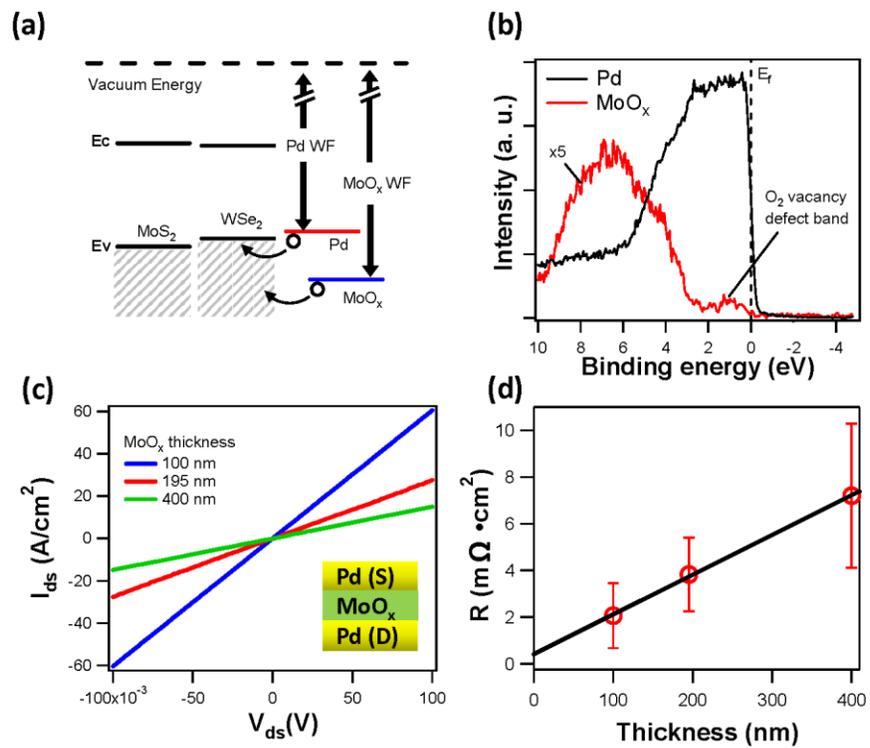

**Figure 2**

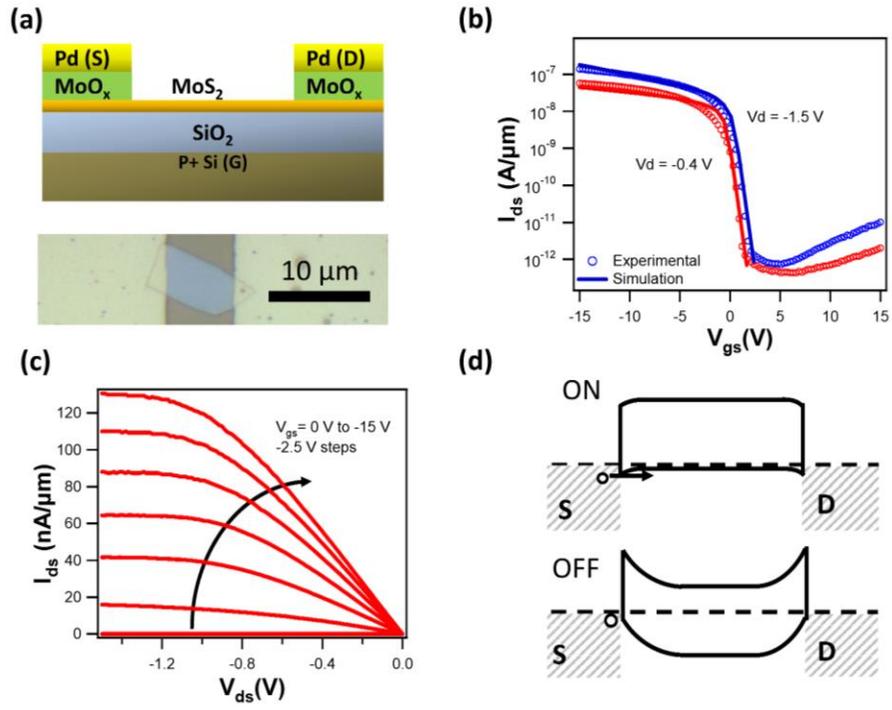

**Figure 3**

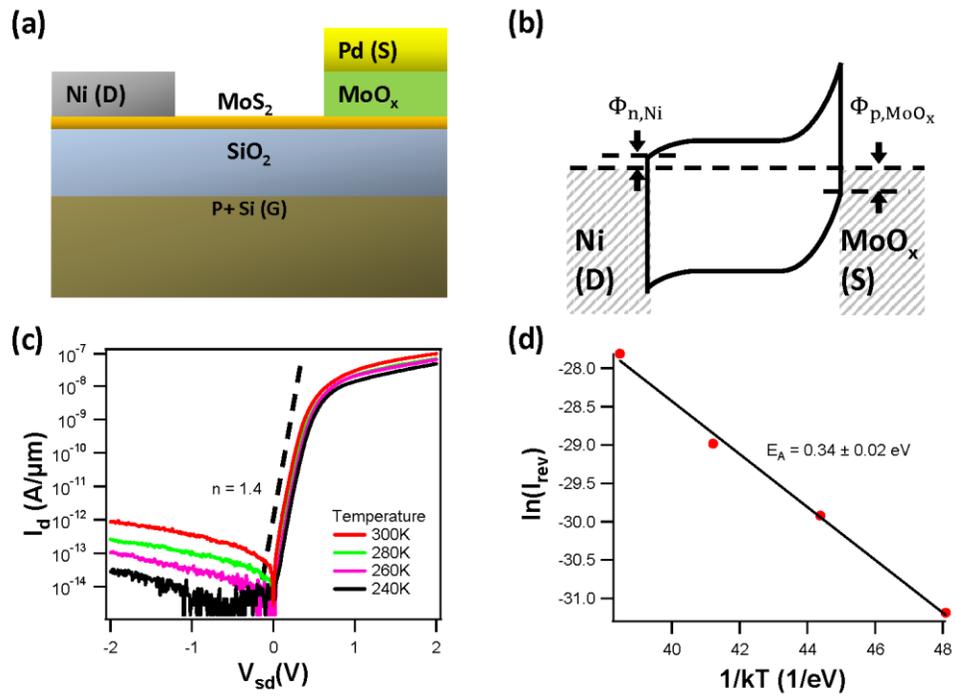

**Figure 4**

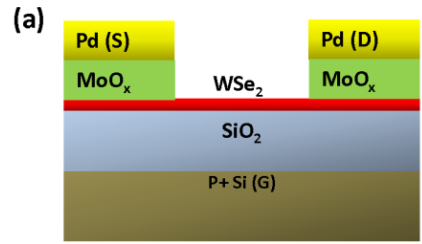
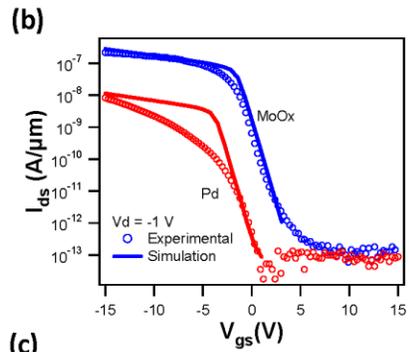
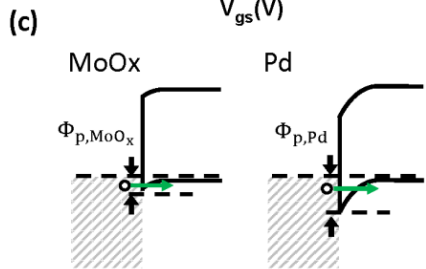

**MoS$_2$ P-type Transistors and Diodes Enabled by High Workfunction MoO$_x$ Contacts**

Steven Chuang[1,2,3], Corsin Battaglia[1,2,3], Angelica Azcatl[4], Stephen McDonnell[4], Jeong Seuk Kang[1,2,3], Xingtian Yin[1,2,3], Mahmut Tosun[1,2,3], Rehan Kapadia[1,2,3], Hui Fang[1,2,3], Robert M. Wallace[4], Ali Javey[1,2,3,*]

[1]*Electrical Engineering and Computer Sciences, University of California, Berkeley, CA, 94720*

[2]*Materials Sciences Division, Lawrence Berkeley National Laboratory, Berkeley, CA 94720*

[3]*Berkeley Sensor and Actuator Center, University of California, Berkeley, CA, 94720*

[4]*Department of Materials Science and Engineering, The University of Texas at Dallas, Richardson, TX, 75080*

* Electronic mail: ajavey@berkeley.edu

# Supplementary Information

## Control MoS$_2$ FETs with Pd and Ni contacts

Figure S1 shows I$_{ds}$ -V$_{gs}$ electrical characteristics of MoS$_2$ FETs with symmetrical a) Pd and b) Ni contacts. The fabrication procedures were the same as the MoO$_x$/MoS$_2$ PFETs, other than the different contact metal used. Specifically, all MoS$_2$ flakes in this study came from the same source crystal. Clear *n*-type characteristics with I$_{on}$/I$_{off}$ > 10$^3$ are exhibited, consistent with literature [1].

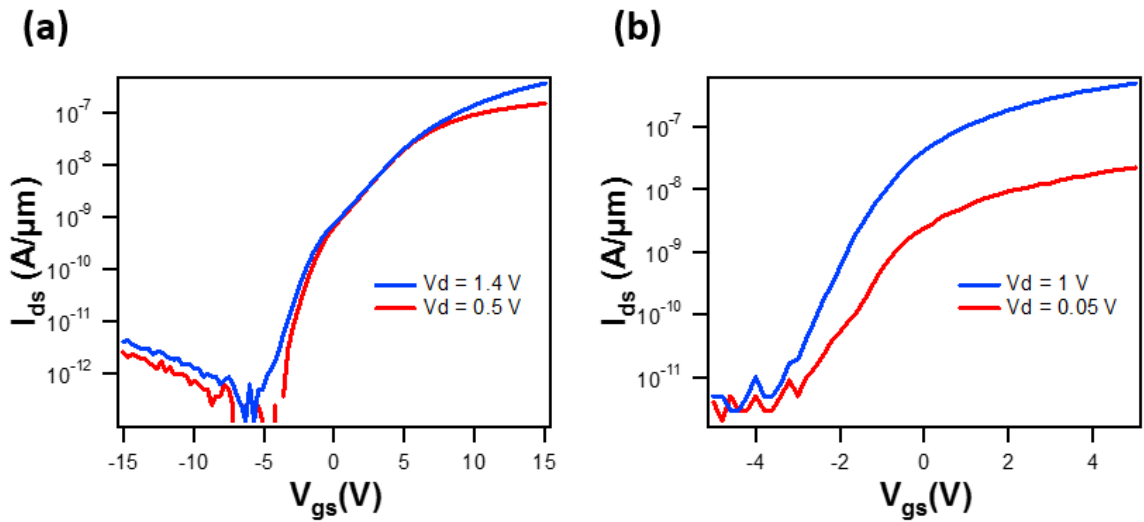

**Figure S1.** I$_{ds}$ vs V$_{gs}$ electrical characteristics of MoS$_2$ PFETs with symmetrical a) Pd and b) Ni contacts.

## Air Stability of $MoS_2$ PFETs with $MoO_x$ contacts

Figure S2a shows the $I_{ds}$-$V_{gs}$ characteristics of a $MoS_2$ PFET with $MoO_x$ contacts measured under vacuum. Both the original measurement and a measurement after >2 weeks exposure to air are plotted. There is no significant difference between the two curves. Figure S2b shows the electrical characteristics of the same device measured in air and in vacuum. Clear degradation in on-current (~ 1 order of magnitude) is shown. This degradation can be attributed to the lowering of $MoO_x$ work function from air exposure previously observed [2]. This degradation is reversible, as the device reverts to its original performance after placement in vacuum.

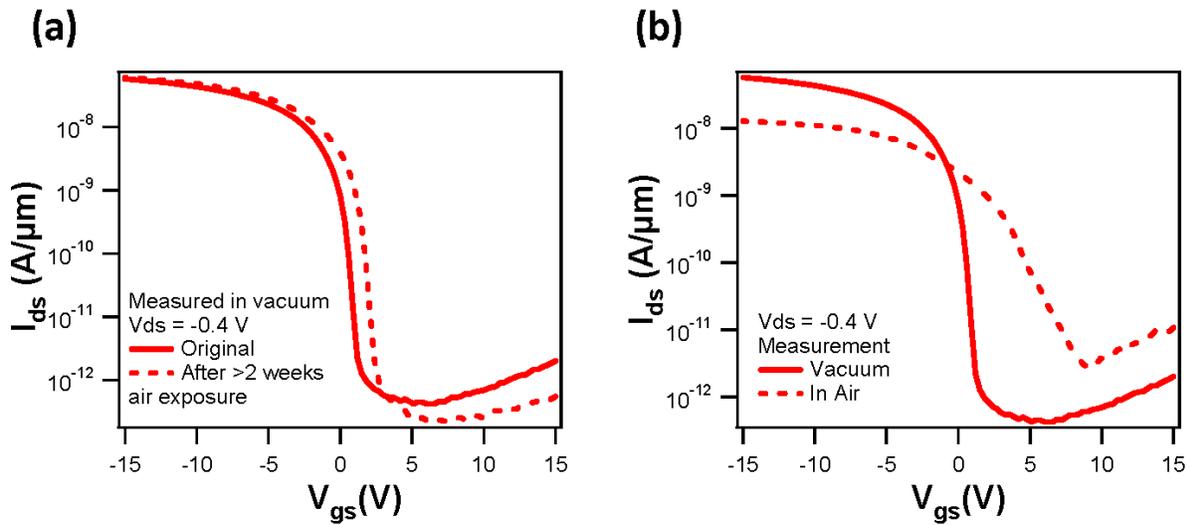

**Figure S2.** $I_{ds}$ - $V_{gs}$ characteristics of a $MoS_2$ PFET with $MoO_x$ contacts a) before and after 2 weeks exposure in air, and b) measured in air and in vacuum.